\documentclass[12pt]{iopart}

\usepackage[colorlinks]{hyperref}

\begin{document}

\title{Focus on the Physics of Cancer}

\author{Thomas Risler$^{1,2,3}$}
\address{$^1$ Institut Curie, PSL Research University, Centre de Recherche, Laboratoire Physico-Chimie Curie UMR 168, 26 rue d'Ulm, F-75248, Paris Cedex 05, France}
\address{$^2$ Sorbonne Universit\'es, UPMC Univ Paris 06, UMR 168, F-75005, Paris, France}
\address{$^3$ CNRS, UMR 168, F-75005, Paris, France}
\ead{thomas.risler@curie.fr}

\begin{abstract}
Despite the spectacular achievements of molecular biology in the second half of the twentieth century and the crucial advances it permitted in cancer research, the fight against cancer has brought some disillusions. It is nowadays more and more apparent that getting a global picture of the very diverse and interlinked aspects of cancer development necessitates, in synergy with these achievements, other perspectives and investigating tools. In this undertaking, multidisciplinary approaches that include quantitative sciences in general and physics in particular play a crucial role. This `focus on' collection contains 19 articles representative of the diversity and state-of-the-art of the contributions that physics can bring to the field of cancer research. 
\end{abstract}


\maketitle

\section{Introduction}

After the discovery of DNA as a the molecule bearing the genetic code and its double-helix structure by Watson and Crick in 1953~\cite{Watson:1953aa}, scientific tools and approaches towards an understanding of the mechanisms at play in living systems started to focus more and more on the molecular level. With the subsequent rise of biochemistry and molecular biology in the 1950s and 1960s, spectacular progress and success were acquired, from identifying individual gene functions and their regulations to understanding complex molecular cascades. The field of cancer research was no exception to this rule. In 1971, President Nixon launched his National Cancer Act, which seeded the molecular research on cancer and the hopes of finding the reductionist principles underlying its tremendous complexity and multiple facets. It was partly fuelled by the publication the year before of a major discovery by two independent groups, namely that of the presence of the reverse transcriptase enzyme in viruses known to trigger tumour formation~\cite{Baltimore:1970aa,Temin:1970aa}. Nixon's War on Cancer, as it came to be called, aimed at identifying the retroviruses thought to underline human cancer. But by the mid-1970s, disillusion came about, as in most cases no human-causing retrovirus could be identified. This marked the beginning of the ups and downs of the modern history of cancer research~\cite{Weinberg:2014aa,mukherjee2010emperor}. The program nevertheless proved to bring major advances. For one, it yielded the discovery of oncogenes and proto-oncogenes, the genes that have the potential to drive malignant cell proliferation either when overexpressed, mutated, or hijacked by retroviruses~\cite{Stehelin:1976aa,Bishop:1983aa}. 

In the subsequent years however, the complexity of the cancer disease at the molecular scale and the impossibility to reduce it to a handful of well-identified molecular events became more and more apparent. First, it was established that cancer development was a multistep process, involving a succession of rate-limiting transformations rather than a single genetic event such as a single-point mutation~\cite{Hahn:1999aa}. Then, in the 1980s and 1990s, the number of identified oncogenes and tumour suppressor genes exploded, and the routes that cancer development could undergo proved to vary depending on the tumour, even from the same organ and even of a given type~\cite{Vogelstein:1989aa,Fearon:1990aa}. These observations underlined the necessity for larger-scale unifying principles to provide a conceptual framework to the myriad of different faces that cancer exhibited. This led to the proposition in the year 2000 of six `classical' hallmarks of cancer, as common biological traits of all human tumours~\cite{Hanahan:2000qf}: self-sufficiency in growth signals, insensitivity to growth-inhibitory signals, evasion of programmed cell death, limitless replicative potential, sustained angiogenesis, and tissue invasion. Two crucial aspects of cancer development were recognized to go along with these hallmarks, namely, genetic diversity and instability for their speed of acquisition and the process of inflammation to support or even promote their function. Along these lines, two years later, some generic principles were also emphasized to underline the processes of invasion and metastases~\cite{Thiery:2002uq}.

But the complexity reappeared again. It indeed soon became apparent that cancer was not only a disease of the cell that can be handled with the tools of molecular biology, but that it had to be addressed globally as a disease of the tumour within its so-called `microenvironment', the environment of apparently normal cells that are recruited during tumourigenesis and which support and potentially promote the cancer hallmarks. Driven by this emerging understanding, 11 years after the publication of the first six hallmarks, the publication of a seventh and a eighth hallmarks---namely, reprogramming of energy metabolism and evading immune destruction---was accompanied by an emphasis of the role of the microenvironment and of its complexity~\cite{Hanahan:2011ly}. Among others, the following important cell types were recognized to participate in tumour development and cancer progression: cancer cells and cancer stem cells, endothelial cells, pericytes, immune inflammatory cells, cancer-associated fibroblasts, and stem and progenitor cells of the tumour stroma. 

Because of this complexity, the initial hope put in Nixon's War on Cancer has been harshly disillusioned. Since the 1950s, age-adjusted cancer mortality rates have declined by only 11\%~\cite{Gravitz:2012aa}, and the prognosis for someone with metastatic cancer is as grim today as it was some 50 years ago. In 2008, {\it Newsweek} ran a story titled ``We Fought Cancer... And Cancer Won''. Facing this reality, scientists from quantitative disciplines such as physicists, mathematicians, computer scientists, and engineers have contributed to cancer research over the years in different ways~\cite{Gravitz:2012aa}. For one obvious contribution, knowledge in physics and further technological developments by engineers have permitted the use of advanced technology in medical imaging and radiation therapy for the diagnosis and treatment of inner tumours. A second contribution is that of bio-informatics in data mining and systems biology, which developed techniques to handle large data sets of genome sequences, gene expression patterns, or metabolic and cell-signalling networks, in their quest for an understanding of the emergent properties of complex biological systems from their multi-agent molecular principles. More recently, a third direction has shown expanding developments, namely, that of seeking a more quantitative understanding of the physical processes involved in the formation of a tumour in its specific microenvironment. Recently, an interagency agreement was created to conduct an international study titled Assessment of PHysical sciences and Engineering advances in LIfe sciences and ONcology (APHELION) (http://wtec.org/aphelion/), led by the National Science Foundation and the National Cancer Institute Office of Physical Sciences – Oncology. The APHELION is aimed at determining the status and trends of research and development, whereby physical sciences and engineering principles are being applied to life sciences, cancer research, and oncology in leading laboratories and organizations in Europe and Asia.
 
In the present review summarizing the `Focus on the Physics of Cancer' published by the {\it New Journal of Physics} over the past three to four years, we shall see different aspects of the contributions and advances that physics can bring to the field of cancer research. The selected pieces of research demonstrate the vast domain of questions and problems that are nowadays addressed by physicists in the field, spanning different scales and tackled with different perspectives, methods, and techniques. Addressed problems include understanding the mechanisms of motility and force production by single cancer cells, the influence of external mechanical constraints on their behaviour (isolated or within a tumour), the importance of the mechanical properties of the tumouric tissue itself, the importance of the spatial structure of the tissue in which the tumour develops as well as of the structures that it creates, the collective migration and the mechanisms of invasion by cancer cells, the evaluation of the efficacy of treatments, and the development of new diagnostic tools. They range from the molecular scale for understanding the mechanisms of force production, motility or response to mechanical perturbations at the single-cell level to multi-cellular or tissue scales when the development of a whole tumour or the influence of its microenvironment are studied. Perspectives and investigating techniques include attempts to study and quantify directly the biological reality by developing either detailed multi-scale models or quantitative experimental techniques for intact {\it in vivo} systems, or, on the contrary, attempts to study a simplified version of the biological reality to extract the essence of a particular question or phenomenon by developing analytical models of idealized situations or studying reconstituted {\it in vitro} experimental systems.

The current collection gathers 19 articles from leading researchers in the field that span the diverse problems, perspectives, and techniques mentioned above. Below, we summarize the main findings of these contributions, grouping them by broad scientific themes. The classification proposed here is somewhat arbitrary, as many bridges, links, and complementary perspectives could have been underlined between these contributions.

\section{Spatial structure}

A first group of contributions to the current focus issue can be identified under the theme of the importance of spatial structure in the development and progression of cancer. These contributions address the role of the spatial organization of the tissue in which a potential tumour develops, the formation of new spatial structures by a population of abnormal cells, as well as the dynamical evolution of a pre-existing structure such as the interface between a proliferative epithelium and its adjacent connective tissue.

Martens {\it et al}~\cite{Martens:2011tg} investigate the influence of the spatial tissue structure in which mutations arise on the dynamics of cancer spread. The standard picture of the acquisition of cancer traits is that mutations that give rise to a proliferative advantage are acquired sequentially over time, leading to a monoclonal population of cells. However, when the pool of pre-cancerous cells is sufficiently spread in space, mutations occur at different locations before a given clone has had the time to spread over the whole population. Therefore, clonal adaptation populations interfere with each other, and the spatial organization of the original cell population influence crucially the global outcome of evolution. The authors apply their model to the case of the human colon, where clonal expansion could be driven by the phenomenon of crypt bifurcation, that is the division of a given crypt into two. Advantageous mutations then spread in the form of waves, as first described by Fisher~\cite{fisher1937wave}, and as has been already recognized to play a potential role for tissues in the context of tumour growth~\cite{ranft2014mechanically}. Because these waves spread with a constant speed, the size of clones grows linearly with time, which is much slower than the characteristic exponential growth observed in well-mixed cell populations. As a consequence, the fixation time of an advantageous mutation is larger, and the waiting time for cancer is increased.

Chatelain {\it et al}~\cite{chatelain2011emergence} investigate the importance of spatial structures from another perspective, namely, that of the structures created by the proliferating cancer cells in an originally homogenous environment. They propose that some characteristic morphological microstructures of skin-cancer lesions such as dots and nests may originate from a phase-separation process between different cell populations. They draw an analogy between the biophysics of tissues and other known phase-separation phenomena such as those observed in block copolymers or in reaction-controlled separating mixtures. Here, segregation occurs when adhesion between cells of the same type becomes large enough, which may correspond to a change in cadherin expressions as observed clinically~\cite{Haass:2005aa}. Microstructures are found to grow in time but eventually saturate due to the control of proliferation by the local nutrient concentration, which leads to the formation of circular patterns. By contrast, asymmetric clusters are observed during the transitory growth regime and resemble those found during early melanoma development. Therefore, the presence of irregular microstructural patterns indicates that the lesion is evolving.

Another example of both the importance of pre-existing spatial structures and the formation of new ones due to cell proliferation is that proposed by Risler and Basan~\cite{Risler:2013aa}, who generalize a previous study on the stability of an interface between a proliferative epithelium and its underlying connective tissue~\cite{Basan:2011fk}. In the epithelium, an oriented flow of cells exists from the basal to the apical side. The originally flat basal interface may become unstable at finite wavelength due to viscous shear stresses caused by the position-dependent flow of cell turnover in the epithelium, which in turn may trigger the formation of fingering protrusions into the connective tissue. An overall increase of cell divisions in the epithelium, which corresponds to higher grades of tumour development, tends to favour this instability. Similarly to the study by Chatelain {\it et al} above, a parallel is drawn with another domain of physics. Here, when the cell-division rate in the epithelium is controlled by the diffusion of nutrients coming from the underlying vascularized tissue, a second wavelength may become unstable, corresponding to a mechanism similar to that of the Mullins-Sekerka instability in the context of diffusion-limited aggregation processes~\cite{mullins1964stability,langer1980instabilities}.

\section{Mechanical influence of the microenvironment and of geometrical confinement on tumour growth and on the development of invasive protrusions}\label{SeModeltumourGrowth}

Another theme present in this collection of papers is that of the importance of the mechanical properties of either the tumour mass or its microenvironment on tumour development. Along these lines and following a previous study~\cite{Montel:2011uq}, Montel {\it et al}~\cite{montel2012isotropic} investigate the effect of an applied mechanical pressure on the long-term growth of a spherical cell aggregate {\it in vitro}, where biochemistry and genetics can be easily controlled. The stress is applied thanks to a difference of osmolarity between the exterior and the interior of the aggregate. They observe that an increase in the applied pressure results on average in a lower cellular proliferation in the spheroid. The applied pressure affects almost exclusively the duplication rate but not substantially the cell-death rate, and more the bulk of the spheroid than the region close to the surface. A comparison with numerical simulations indicates that these observations can originate from pure mechanical effects.

Drasdo and Hoehme~\cite{drasdo2012modeling} explore numerically the potential biomechanical influences of cell-cycle entrance and cell migration on the growth and invasion pattern of a growing tumour, depending on its surrounding environment. In a free environment, cells may exert passive pushing forces coming from cellular proliferation (as is the case in the model proposed by Risler and Basan~\cite{Risler:2013aa}) as well as pulling forces due to the stretching of cell-cell contacts. The authors attempt to decipher between the patterns resulting from these two distinct mechanisms. Putting an environment of granular objects and cells of another type to mimic the effect of embedding tissues, the authors find the potential formation of invasive fingers, which become particularly pronounced when friction between the cells and the substrate of the embedding tissue is increased. The authors' results resemble those observed experimentally for the growth dynamics of multi-cellular spheroids in agarose gels~\cite{Helmlinger:1997if}.

Another study that investigates the dynamics of the growth of a whole tumour is that proposed by Scium\`e {\it et al}~\cite{Sciume:2013aa}. The authors present a multiphase model for the growth of a tumour mass composed of extracellular matrix, tumour cells, healthy cells and an interstitial fluid. The interstitial fluid helps the transport of nutrients, and tumour cells can become necrotic upon exposure to low nutrient concentrations or excessive mechanical pressure. The equations are solved by finite-element methods in three different cases: 1) First that of a multicellular spheroid in a culture medium, where the authors validate the model by experimental data. 2) Second that of a spheroid within a healthy tissue and extracellular matrix, where they observe a reduced growth rate as compared to the previous case. Tumour cells may eventually either displace the healthy cells or infiltrate the healthy tissue, depending on the relative adhesion of the two cell populations to the extracellular matrix. 3) Third that of a tumour cord where the malignant cells grow around microvessels. Similarities in the mechanical importance of pressure can be drawn with the experimental study of Montel {\it et al}~\cite{montel2012isotropic}, and the potential patterns of tumour growth or invasion can be compared with the numerical study of Drasdo and Hoehme~\cite{drasdo2012modeling}.

\section{Statistical models of collective cell migration and tumour spreading}

In tumour spreading and metastases' dissemination, collective cell migration potentially plays a crucial role~\cite{Friedl:2009fk}. Nnetu {\it et al}~\cite{nnetu2012impact} challenge the standard picture that collective cell migration relies mostly on cellular interactions such as cell-cell adhesions. Instead, they propose that jamming and glass-like effects play an important role in keeping the population cohesive, even in the absence of cell-cell adhesion. Using malignant and non-malignant epithelial cell lines as well as fibroblasts in a two-dimensional migration assay, the authors find that in the core of a propagating front and as a result of jamming, cells move ballistically rather than randomly. As a result, when a cell escapes and starts moving randomly, the propagating front catches up such that an effective stable boundary is observed on long time scales. When two monolayers meet, slowed-down dynamics and jamming effects lead to the formation of stable borders, even between monolayers of the same cell type.

Lee {\it et al}~\cite{Lee:2013aa} characterize the flow field of a migrating sheet of cells acquired with particle-image velocimetry. Using a finite-time Lyapunov exponent analysis, the authors find that the flow field is not chaotic. Stretching of the sheet is localized at the leading edge of migration and increases with stimulation. Surprisingly, they also find that plastic rearrangements increase with increasing cell densities, an observation that is in contrast with inanimate systems.

Along these lines but modelling an {\it in-vivo} three-dimensional situation, Fort and Sol\'e~\cite{fort2013accelerated} develop a model of biased random walk to investigate the cell dispersal of gliobastomas, which are highly diffuse, malignant tumours. The authors show the importance of the bias effect from adjusting the values of three relevant parameters of an analytical model, without the need to perform heavy numerical simulations. Potentially, the authors' model could allow incorporation of the bias effect into the simulations of glioblastoma invasion that are nowadays currently performed for individual patients~\cite{Suarez:2012aa}. This could grandly modify the predictions of such simulations and, therefore, lead to potentially essential clinical implications.

\section{Cytoskeleton and sub-cellular processes - force generation at the single-cell level}\label{SeSubcellular}

Four contributions of the focus issue investigate the production of the mechanical forces relevant for metastatic cell invasion and motility at the single-cell level. Cell motility relies on a balance of biomechanical processes such as cell adhesion and de-adhesion~\cite{Friedl:2000aa}, cytoskeletal remodelling~\cite{Mierke:2011aa}, protrusive force generation~\cite{Friedl:2000aa}, as well as on matrix properties such as stiffness, pore size, protein composition and enzymatic degradation~\cite{Zaman:2006aa}. In this context, Mierke~\cite{mierke2013integrin} investigates the role the $\alpha$v$\beta$3 integrin in cancer-cell invasion through increased cellular stiffness and enhanced cytoskeletal remodelling, which allow the generation and transmission of contractile forces. Using different cancer cell lines expressing high or low levels of this integrin (respectively $\alpha$v$\beta$3$^{\rm high}$ and $\alpha$v$\beta$3$^{\rm low}$ cells), she finds that $\alpha$v$\beta$3$^{\rm high}$ cells have a threefold increase in their invasion compared to $\alpha$v$\beta$3$^{\rm low}$ cells. Using a myosin light chain kinase inhibitor, she could show that contractile forces are essential for the increase in cellular stiffness mediated by $\alpha$v$\beta$3 integrins and subsequently for the enhanced cancer-cell invasion. She could, for example, rescue the cellular invasiveness of $\alpha$v$\beta$3$^{\rm low}$ cells after addition of the contractility enhancer calyculin A.

Kristal-Muscal {\it et al}~\cite{kristal2013metastatic} study the process of substrate indentation by metastatic cells, corresponding to the initial stage of metastatic penetration into the adjacent tissue and extracellular matrix. Thanks to an {\it in vitro} system, they mimic the penetration of metastatic cells into the extracellular matrix or through the endothelial cells of a blood-vessel wall. Their observations reveal the existence of an indentation process based purely on the production of mechanical forces. Cells with higher metastatic potential are softer but are found to apply stronger forces than non-metastatic cells, and benign cells do not indent substrates at all. Cells are also found to develop forces more readily on stiffer gels, which provide grip handles for the cells to hold on to. Therefore, substrate stiffness and adhesion properties may be viable targets against metastatic penetration, suggesting new potential avenues of treatment.

After indentation and penetration of the cancer cells into adjacent tissues, cancer-cell motility---their ability to move---is thought to play an important role in the colonization of other tissues and the formation of metastases~\cite{Friedl:2004aa,wedlich2006cell,Risler:2009rp}. Along these lines, two papers of this series investigate the mechanisms of cell motility at the level of the cell lamellipodium. Zimmermann and Falcke~\cite{zimmermann2013existence} model the lamellipodium as a viscoelastic actin gel in its bulk and a dynamic boundary layer of newly polymerized filaments at its leading edge. The authors find three different parameter regimes: a stable, stationarily protruding lamellipodium; a stable lamellipodium showing oscillatory motion of the leading edge; and no stable lamellipodium with zero filament density. The authors also investigate the dynamic force-velocity relation and predict that it should change if the cell experiences a constant force for a long time. This is due to the fact that filament number can adjust if a force is applied for a long time and should therefore happen independently of cell signalling~\cite{Carlsson:2003aa,Parekh:2005aa}. The authors point out that since cells typically experience the forces exerted by surrounding tissues over a long time, dynamic force-velocity relations as traditionally measured {\it in vitro} may not be relevant for actual tissues.

Another study that addresses the mechanisms of cell motility and lamellipodium mechanics is that by Havrylenko {\it et al}~\cite{Havrylenko:2014aa}, who investigate a mechanism of cell motility that does not rely on an acto-myosin cytoskeleton, as does cell migration in mammalian systems. Their model system is the {\it Caenorhabditis elegans} sperm cell, which has a cytoskeleton that is biochemically different and has no structural similarity to actin, and for which no associated molecular motors is known~\cite{Nelson:1982aa,Roberts:2000aa}. Studying the migration properties of these cells in different adhesive conditions, the authors find that their migration nevertheless displays the same molecular characteristics as that of acto-myosin-containing cells. In particular, they find the existence of a backward (retrograde) flow of the cytoskeleton toward the cell body, resulting from slippage of the cytoskeleton with respect to the substrate and driven by cytoskeletal assembly and contractility.

\section{Influence of external biomechanical conditions on mechanosensing and on mechanical regulation of cancer cells and tissues}

A substantial number of papers of this series look at the importance of the biomechanical properties of cancer cells, tumouric tissues, and their environments, as well as on the influence of external forces on these or other properties, including cell fate. This is the case of papers already discussed above, such as those presented in sections~\ref{SeModeltumourGrowth} and \ref{SeSubcellular}, for example. But external biomechanical conditions can also have a direct influence at the cell level on properties as essential as cell survival. The study by Montel {\it et al}~\cite{montel2012isotropic} discussed in section~\ref{SeModeltumourGrowth} shows such an example. Another example is that studied by Mitchell and King~\cite{Mitchell:2013aa}, who investigate the influence of hydrodynamic shear forces on the survival of circulating tumour cells. In particular, the sensitization of colon and prostate cancer cells to apoptotic agents by fluid shear forces increases in an intensity- and time-dependent manner. Low interstitial fluid flows such as those found in the tumour microenvironment are not sufficient to influence the cell response but the flows present in the blood circulation could explain why only a small portion of circulating tumour cells survive and generate metastases~\cite{Fidler:2002aa}.

Nolting and K\"oster~\cite{nolting2013influence} study another aspect of the influence of fluid shear flows, namely, that on the networks of keratin intermediate filaments. Intermediate filaments are part of the cytoskeleton along with microtubules, actin filaments, associated proteins, and molecular motors~\cite{brayB,Pollard:2003lr,Risler:2009rp}. One function of these fibrous proteins is to withstand potentially harmful mechanical influences and to guarantee the integrity of the cell~\cite{Fletcher:2010aa,Herrmann:2004aa}. As it was previously known that the dynamics of the keratin network is influenced when the cell is exposed to shear stresses, it was unknown which part of the dynamics and by which mechanism the keratin network is affected. Upon application of fluid shear flow, the authors find that bundle dynamics is reduced on a timescale of minutes. They show evidence that the regulation is active and comes from the acto-myosin network, which rigidifies and transmits its rigidity to the keratin network. Therefore, the cytoskeletal cross-talk between keratin and actin networks appears to be shear-stress dependent.

Two papers of the focus series study the influence of either the microenvironment or the external mechanical constraints on the stiffness properties of individual cells or entire tissues. It is indeed now established in several tumour types that individual tumour cells have distinct biophysical and biomechanical properties as compared to their healthy counterparts~\cite{fritsch2010biomechanical,PhysSciOncCentNetwork:2013aa}, and that this is the case also at the tissue level (see below). Using an Atomic Force Microscope with a 5.3 $\mu$m diameter spherical probe, Guo {\it et al}~\cite{guo2014effect} study the stiffness of individual human mammary epithelial cells in four different phases of cancer progression, namely, studying normal (non-transformed), immortal, tumourigenic, and metastatic cells. They determine the elastic moduli of cells in all four phases as a function of the subregion of the cell (over the nucleus versus over the cytoplasm) and of the cell's microenvironment (inside, at the periphery, or isolated outside of a contiguous cellular monolayer). The authors find that there are only minor to negligible differences in stiffness between cellular subregions and that cells become globally softer as they advance to the tumourigenic phase, except in the final step to becoming metastatic. They find that normal epithelial cells are stiffer when surrounded by other cells within a monolayer but that the microenvironment has only a slight effect on transformed cells, sometimes opposite.

At the tissue level, a common feature of several disease states, like fibrosis or some types of cancers such as breast, colon or pancreatic cancers, is distinct biophysical properties from those of the normal tissues from which they originate. For example, diseased tissues often present an increased interstitial fluid pressure~\cite{Boucher:1997aa} and solid tissue stress~\cite{Stylianopoulos:2013aa,Basan:2009vn}. They are also often stiffer~\cite{Samuel:2011aa}, a fact that has been suggested to contribute to disease progression~\cite{Schrader:2011aa}. Pogoda {\it et al}~\cite{pogoda2014compression} study the elastic properties of normal and tumouric brain tissues. The authors show that normal-brain and glioma tissues increase their shear elastic moduli under modest uniaxial compression but not under elongation or increased shear strains, the effect being more pronounced for glioma tissue. It is suggested that compression stiffening, which might occur with the increased vascularization and interstitial pressure gradients that are characteristic of glioma and other solid tumours, effectively stiffens the environment. From the {\it in vitro} observations regarding the response of glioma cells to substrate stiffness change~\cite{Georges:2006qf,Flanagan:2002aa}, the increased local stiffness might contribute to increased tension, motility and proliferation of the tumour cells.

\section{Modelling of treatments and imaging techniques}

The contribution of physics to the field of cancer research presented in this `focus on' series does not end with the attempt at understanding the properties of cancer cells, tissues and disease progression. It is also shown that physics contributes in modelling the efficacy and principles of diagnostic techniques and of treatment delivery. Two examples of the contribution of physics in these domains are presented in the current series. Following on the specificities of the biomechanical properties of tumouric tissues, Simon {\it et al}~\cite{simon2013non} investigate magnetic resonance elastography (MRE) as a tool for the clinical diagnosis of intracranial neoplasm. MRE is a non-invasive medical imaging technique that measures the mechanical properties of soft tissues by introducing shear waves and imaging their propagation, thanks to magnetic resonance imaging (MRI)~\cite{Muthupillai:1995aa}. It is used in a variety of diseases' diagnoses, but MRE still suffers from limited spatial resolution due to ill-posed inverse problems required for parameter recovery~\cite{Manduca:2001aa}. In their work, the authors improved the capability of MRE to obtain spatially resolved maps of viscoelastic constants in the biomechanical characterization of cerebral tumours in their natural environment. Their preliminary data reveal a loss of stiffness in malignancies compared to healthy reference tissues or benign variants. It should be noted that this is somewhat in contradiction with the study of Pogoda {\it et al}~\cite{pogoda2014compression} discussed above, which predicts a stiffening of glioma tissues due to compression. It might be that Simon {\it et al} find the opposite because of different tumour types, disease stage or degree of compression. The softening of numbers of tumours may be due to a reduction in cross-linking network capability or structure, a notion supported by biophysical single-cell studies. Cancerous behaviour of tumour cells has indeed been attributed to cytoskeletal transformations, which soften the cell's response to small deformations and potentially increases its invasive aggressiveness~\cite{fritsch2010biomechanical} or facilitates the tumour growth at the extent of the neighbouring tissue~\cite{Katira:2012oq}. When sufficiently sensitive, MRE could provide a predictive marker for tumour malignancy and thereby contribute to an early non-invasive clinical assessment of suspicious cerebral lesions.

The second presented contribution to improving the efficacy of cancer treatment is that of modelling the delivery of nanotherapeutics, which aim at targeting specifically the diseased tissue. Nanotherapeutics consists in concentrating the drugs inside nanoscale vehicles, in order to enhance drug accumulation within tumours as compared with conventional chemotherapeutics, which relies on pure passive diffusion. In this series, van de Ven {\it et al}~\cite{Ven:2013aa} develop a theoretical framework for modelling the delivery of nanotherapeutics based on the automated evaluation of vascular perfusion curves measured at the single-vessel level. The vascular perfusion curves contain data acquired using video-rate laser-scanning microscopy and consisting of blood flow velocity, flux and hematocrit measured by tracking trajectories of fluorescent red blood cells~\cite{Kamoun:2010aa}. This approach enables an automated ranking of tumour vascular perfusion in order to model the delivery of nanotherapeutics, without requiring any underlying assumptions about tissue structure, function, and heterogeneities. Such rankings can be correlated with a variety of quantifiable physiological parameters in order to evaluate the behaviour of a given tumour. The resulting rankings are found to correlate inversely with experimental nanoparticle accumulation measurements. With additional calibration, these methodologies may enable the investigation of nanotherapeutics delivery strategies in a variety of tumour models.

\section{Summary and outlook}

The `Focus on the Physics of Cancer' collection offers various examples of the state of the art of the contribution of physical sciences to the field of cancer research. These span different scales ranging from molecular assemblies to tissue spatial organization. They present a large variety of different questions such as the role of biomechanics in tumour growth and cancer-cell behaviour, the role of the microenvironment and of its biophysical properties, the role and mechanisms of the collective migration of cancer cells, and the cytoskeletal organization at the single-cell level and its influence on cell mechanics and migration, as well as the exploitation of particular biophysical properties of tissues and of the circulating system in developing new diagnostic tools and drug delivery strategies.

The history of cancer research has been paved with ups and downs, with reductionist hopes and periods of harsh disillusions, demonstrating the need of a wider-angle of view than molecular biology alone would be able to offer. Cross-disciplinary collaborations have had a long history in cancer research. They enable to grasp the mechanisms of the cancer disease at greater spatial and temporal scales and with wider perspectives. In this collective undertaking, physics contributes in a crucial way.


\section*{References}

\begin{thebibliography}{10}

\bibitem{Watson:1953aa}
J~D Watson and F~H Crick.
\newblock Molecular structure of nucleic acids: a structure for deoxyribose
  nucleic acid.
\newblock {\em Nature}, 171(4356):737--8, 1953.

\bibitem{Baltimore:1970aa}
D~Baltimore.
\newblock {RNA}-dependent {DNA} polymerase in virions of {RNA} tumour viruses.
\newblock {\em Nature}, 226(5252):1209--11, 1970.

\bibitem{Temin:1970aa}
H~M Temin and S~Mizutani.
\newblock {RNA}-dependent {DNA} polymerase in virions of {R}ous sarcoma virus.
\newblock {\em Nature}, 226(5252):1211--3, 1970.

\bibitem{Weinberg:2014aa}
R~A Weinberg.
\newblock Coming full circle-from endless complexity to simplicity and back
  again.
\newblock {\em Cell}, 157(1):267--71, 2014.

\bibitem{mukherjee2010emperor}
S Mukherjee.
\newblock {\em The Emperor of all Maladies: a Biography of Cancer}.
\newblock (New York: Simon and Schuster), 2010.

\bibitem{Stehelin:1976aa}
D~Stehelin, H~E Varmus, J~M Bishop, and P~K Vogt.
\newblock {DNA} related to the transforming gene(s) of avian sarcoma viruses is
  present in normal avian {DNA}.
\newblock {\em Nature}, 260(5547):170--3, 1976.

\bibitem{Bishop:1983aa}
J~M Bishop.
\newblock Cellular oncogenes and retroviruses.
\newblock {\em Annu. Rev. Biochem.}, 52:301--54, 1983.

\bibitem{Hahn:1999aa}
W~C Hahn, C~M Counter, A~S Lundberg, R~L Beijersbergen, M~W Brooks, and R~A
  Weinberg.
\newblock Creation of human tumour cells with defined genetic elements.
\newblock {\em Nature}, 400(6743):464--8, 1999.

\bibitem{Vogelstein:1989aa}
B~Vogelstein, E~R Fearon, S~E Kern, S~R Hamilton, A~C Preisinger, Y~Nakamura,
  and R~White.
\newblock Allelotype of colorectal carcinomas.
\newblock {\em Science}, 244(4901):207--11, 1989.

\bibitem{Fearon:1990aa}
E~R Fearon and B~Vogelstein.
\newblock A genetic model for colorectal tumorigenesis.
\newblock {\em Cell}, 61(5):759--67, 1990.

\bibitem{Hanahan:2000qf}
D~Hanahan and R~A Weinberg.
\newblock The hallmarks of cancer.
\newblock {\em Cell}, 100(1):57--70, 2000.

\bibitem{Thiery:2002uq}
J-P Thiery.
\newblock Epithelial-mesenchymal transitions in tumour progression.
\newblock {\em Nat. Rev. Cancer}, 2(6):442--54, 2002.

\bibitem{Hanahan:2011ly}
D Hanahan and R~A Weinberg.
\newblock Hallmarks of cancer: the next generation.
\newblock {\em Cell}, 144(5):646--74, 2011.

\bibitem{Gravitz:2012aa}
L Gravitz.
\newblock Physical scientists take on cancer.
\newblock {\em Nature}, 491(7425):S49, 2012.

\bibitem{Martens:2011tg}
E~A Martens, R Kostadinov, C~C Maley, and O Hallatschek.
\newblock Spatial structure increases the waiting time for cancer.
\newblock {\em New J. Phys.}, 13(11):115014, 2011.

\bibitem{fisher1937wave}
R~A Fisher.
\newblock The wave of advance of advantageous genes.
\newblock {\em Annals of Eugenics}, 7(4):355--369, 1937.

\bibitem{ranft2014mechanically}
J Ranft, M Aliee, J Prost, F J{\"u}licher, and
  J-F Joanny.
\newblock Mechanically driven interface propagation in biological tissues.
\newblock {\em New J. Phys.}, 16(3):035002, 2014.

\bibitem{chatelain2011emergence}
C~Chatelain, T~Balois, P~Ciarletta, and M~Ben Amar.
\newblock Emergence of microstructural patterns in skin cancer: a phase
  separation analysis in a binary mixture.
\newblock {\em New J. Phys.}, 13(11):115013, 2011.

\bibitem{Haass:2005aa}
N~K Haass and M Herlyn.
\newblock Normal human melanocyte homeostasis as a paradigm for understanding
  melanoma.
\newblock {\em J. Investig. Dermatol. Symp. Proc.}, 10(2):153--63, 2005.

\bibitem{Risler:2013aa}
T Risler and M Basan.
\newblock Morphological instabilities of stratified epithelia: a mechanical
  instability in tumour formation.
\newblock {\em New J. Phys.}, 15(6):065011, 2013.

\bibitem{Basan:2011fk}
M Basan, J-F Joanny, J Prost, and T Risler.
\newblock Undulation instability of epithelial tissues.
\newblock {\em Phys. Rev. Lett.}, 106(15):158101, 2011.

\bibitem{mullins1964stability}
W~W Mullins and R~F~Sekerka.
\newblock Stability of a planar interface during solidification of a dilute
  binary alloy.
\newblock {\em J. Appl. Phys.}, 35(2):444--451, 1964.

\bibitem{langer1980instabilities}
J~S~Langer.
\newblock Instabilities and pattern formation in crystal growth.
\newblock {\em Rev. Mod. Phys.}, 52(1):1, 1980.

\bibitem{Montel:2011uq}
F Montel {\it et al}.
\newblock Stress clamp experiments on multicellular tumor spheroids.
\newblock {\em Phys. Rev. Lett.}, 107(18):188102, 2011.

\bibitem{montel2012isotropic}
F~Montel, M~Delarue, J~Elgeti, D~Vignjevic, G~Cappello, and J~Prost.
\newblock Isotropic stress reduces cell proliferation in tumor spheroids.
\newblock {\em New J. Phys.}, 14(5):055008, 2012.

\bibitem{drasdo2012modeling}
D~Drasdo and S~Hoehme.
\newblock Modeling the impact of granular embedding media, and pulling versus
  pushing cells on growing cell clones.
\newblock {\em New J. Phys.}, 14(5):055025, 2012.

\bibitem{Helmlinger:1997if}
G~Helmlinger, P~A Netti, H~C Lichtenbeld, R~J Melder, and R~K Jain.
\newblock Solid stress inhibits the growth of multicellular tumor spheroids.
\newblock {\em Nat. Biotechnol.}, 15(8):778--83, 1997.

\bibitem{Sciume:2013aa}
G~Scium{\`e}, S~Shelton, W~Gray, C~Miller, F~Hussain, M~Ferrari, P~Decuzzi,
  and B~Schrefler.
\newblock A multiphase model for three-dimensional tumor growth.
\newblock {\em New J. Phys.}, 15:015005, 2013.

\bibitem{Friedl:2009fk}
P Friedl and D Gilmour.
\newblock Collective cell migration in morphogenesis, regeneration and cancer.
\newblock {\em Nat. Rev. Mol. Cell Biol.}, 10(7):445--57, 2009.

\bibitem{nnetu2012impact}
K~D Nnetu, M Knorr, J K{\"a}s, and M Zink.
\newblock The impact of jamming on boundaries of collectively moving
  weak-interacting cells.
\newblock {\em New J. Phys.}, 14(11):115012, 2012.

\bibitem{Lee:2013aa}
R~M Lee, D~H Kelley, K~N Nordstrom, N~T Ouellette, and
  W Losert.
\newblock Quantifying stretching and rearrangement in epithelial sheet
  migration.
\newblock {\em New J. Phys.}, 15(2):025036, 2013.

\bibitem{fort2013accelerated}
J Fort and R~V Sole.
\newblock Accelerated tumor invasion under non-isotropic cell dispersal in
  glioblastomas.
\newblock {\em New J. Phys.}, 15(5):055001, 2013.

\bibitem{Suarez:2012aa}
C Suarez, F Maglietti, M Colonna, K Breitburd, and
  G Marshall.
\newblock Mathematical modeling of human glioma growth based on brain
  topological structures: study of two clinical cases.
\newblock {\em PLoS One}, 7(6):e39616, 2012.

\bibitem{Friedl:2000aa}
P~Friedl and E~B Br{\"o}cker.
\newblock The biology of cell locomotion within three-dimensional extracellular
  matrix.
\newblock {\em Cell. Mol. Life Sci.}, 57(1):41--64, 2000.

\bibitem{Mierke:2011aa}
C~T Mierke.
\newblock Cancer cells regulate biomechanical properties of human microvascular
  endothelial cells.
\newblock {\em J. Biol. Chem.}, 286(46):40025--37, 2011.

\bibitem{Zaman:2006aa}
M~H Zaman, L~M Trapani, A~L Sieminski, A Siemeski, D
  Mackellar, H Gong, R~D Kamm, A Wells, D~A Lauffenburger,
  and P Matsudaira.
\newblock Migration of tumor cells in 3{D} matrices is governed by matrix
  stiffness along with cell-matrix adhesion and proteolysis.
\newblock {\em Proc. Natl. Acad. Sci. USA}, 103(29):10889--94, 2006.

\bibitem{mierke2013integrin}
C~T Mierke.
\newblock The integrin alphav beta3 increases cellular stiffness and
  cytoskeletal remodeling dynamics to facilitate cancer cell invasion.
\newblock {\em New J. Phys.}, 15(1):015003, 2013.

\bibitem{kristal2013metastatic}
R~Kristal-Muscal, L Dvir, and D Weihs.
\newblock Metastatic cancer cells tenaciously indent impenetrable, soft
  substrates.
\newblock {\em New J. Phys.}, 15(3):035022, 2013.

\bibitem{Friedl:2004aa}
P Friedl, Y Hegerfeldt, and M Tusch.
\newblock Collective cell migration in morphogenesis and cancer.
\newblock {\em Int. J. Dev. Biol.}, 48(5-6):441--9, 2004.

\bibitem{wedlich2006cell}
D Wedlich.
\newblock {\em Cell migration in development and disease}.
\newblock (New York: Wiley), 2006.

\bibitem{Risler:2009rp}
T Risler.
\newblock Cytoskeleton and Cell Motility.
\newblock In Robert~A. Meyers, editor, {\em Encyclopedia of Complexity and
  Systems Science}, volume Part 3, pages 1738--74. (New York: Springer), 2009.

\bibitem{zimmermann2013existence}
J Zimmermann and M Falcke.
\newblock On the existence and strength of stable membrane protrusions.
\newblock {\em New J. Phys.}, 15(1):015021, 2013.

\bibitem{Carlsson:2003aa}
A~E Carlsson.
\newblock Growth velocities of branched actin networks.
\newblock {\em Biophys. J.}, 84(5):2907--18, 2003.

\bibitem{Parekh:2005aa}
S~H Parekh, O Chaudhuri, J~A Theriot, and D~A Fletcher.
\newblock Loading history determines the velocity of actin-network growth.
\newblock {\em Nat. Cell Biol.}, 7(12):1219--23, 2005.

\bibitem{Havrylenko:2014aa}
S Havrylenko, X Mezanges, E Batchelder, and J Plastino.
\newblock Extending the molecular clutch beyond actin-based cell motility.
\newblock {\em New J. Phys.}, 16(10), 2014.

\bibitem{Nelson:1982aa}
G~A Nelson, T~M Roberts, and S~Ward.
\newblock Caenorhabditis elegans spermatozoan locomotion: amoeboid movement
  with almost no actin.
\newblock {\em J. Cell Biol.}, 92(1):121--31, 1982.

\bibitem{Roberts:2000aa}
T~M Roberts and M~Stewart.
\newblock Acting like actin: the dynamics of the nematode major sperm protein
  ({MSP}) cytoskeleton indicate a push-pull mechanism for amoeboid cell
  motility.
\newblock {\em J. Cell Biol.}, 149(1):7--12, 2000.

\bibitem{Mitchell:2013aa}
M~J Mitchell and M~R King.
\newblock Fluid shear stress sensitizes cancer cells to receptor-mediated
  apoptosis via trimeric death receptors.
\newblock {\em New J. Phys.}, 15:015008, 2013.

\bibitem{Fidler:2002aa}
I~J Fidler, S Yano, R-D Zhang, T Fujimaki, and C~D
  Bucana.
\newblock The seed and soil hypothesis: vascularisation and brain metastases.
\newblock {\em Lancet Oncol.}, 3(1):53--7, 2002.

\bibitem{nolting2013influence}
J-F Nolting and S K{\"o}ster.
\newblock Influence of microfluidic shear on keratin networks in living cells.
\newblock {\em New J. Phys.}, 15(4):045025, 2013.

\bibitem{brayB}
D Bray.
\newblock {\em {Cell Movements}}.
\newblock 2nd edn (New York: Garland), 2001.

\bibitem{Pollard:2003lr}
T~D Pollard.
\newblock The cytoskeleton, cellular motility and the reductionist agenda.
\newblock {\em Nature}, 422(6933):741--5, 2003.

\bibitem{Fletcher:2010aa}
D~A Fletcher and R~D Mullins.
\newblock Cell mechanics and the cytoskeleton.
\newblock {\em Nature}, 463(7280):485--92, 2010.

\bibitem{Herrmann:2004aa}
H Herrmann and U Aebi.
\newblock Intermediate filaments: molecular structure, assembly mechanism,
  and integration into functionally distinct intracellular scaffolds.
\newblock {\em Annu. Rev. Biochem.}, 73:749--89, 2004.

\bibitem{fritsch2010biomechanical}
A~Fritsch, M~H{\"o}ckel, T~Kiessling, K~D Nnetu, F~Wetzel, M~Zink, and
  J~A K{\"a}s.
\newblock Are biomechanical changes necessary for tumour progression?
\newblock {\em Nature Phys.}, 6(10):730--732, 2010.

\bibitem{PhysSciOncCentNetwork:2013aa}
D~B Agus {\it et al} (Physical Sciences - Oncology Centers Network).
\newblock A physical sciences network characterization of non-tumorigenic and
  metastatic cells.
\newblock {\em Sci. Rep.}, 3:1449, 2013.

\bibitem{guo2014effect}
X Guo, K Bonin, K Scarpinato, and M Guthold.
\newblock The effect of neighboring cells on the stiffness of cancerous and
  non-cancerous human mammary epithelial cells.
\newblock {\em New J. Phys.}, 16(10):105002, 2014.

\bibitem{Boucher:1997aa}
Y~Boucher, H~Salehi, B~Witwer, G~R Harsh $\rm I\!V$, and R~K Jain.
\newblock Interstitial fluid pressure in intracranial tumours in patients and
  in rodents.
\newblock {\em Br. J. Cancer}, 75(6):829--36, 1997.

\bibitem{Stylianopoulos:2013aa}
T Stylianopoulos, J~D Martin, M Snuderl, F Mpekris,
  S~R Jain, and R~K Jain.
\newblock Coevolution of solid stress and interstitial fluid pressure in tumors
  during progression: implications for vascular collapse.
\newblock {\em Cancer Res.}, 73(13):3833--41, 2013.

\bibitem{Basan:2009vn}
M Basan, T Risler, J-F Joanny, X Sastre-Garau,
  and J Prost.
\newblock Homeostatic competition drives tumor growth and metastasis
  nucleation.
\newblock {\em HFSP J.}, 3(4):265--72, 2009.

\bibitem{Samuel:2011aa}
M~S Samuel {\it et al}.
\newblock Actomyosin-mediated cellular tension drives increased tissue
  stiffness and beta-catenin activation to induce epidermal hyperplasia and
  tumor growth.
\newblock {\em Cancer Cell}, 19(6):776--91, 2011.

\bibitem{Schrader:2011aa}
J Schrader, T~T Gordon-Walker, R~L Aucott, M van
  Deemter, A Quaas, S Walsh, D Benten, S~J Forbes,
  R~G Wells, and J~P Iredale.
\newblock Matrix stiffness modulates proliferation, chemotherapeutic response,
  and dormancy in hepatocellular carcinoma cells.
\newblock {\em Hepatology}, 53(4):1192--205, 2011.

\bibitem{pogoda2014compression}
K Pogoda, L Chin, P~C Georges, F~J Byfield, R
  Bucki, R Kim, M Weaver, R~G Wells, C Marcinkiewicz,
  and P~A Janmey.
\newblock Compression stiffening of brain and its effect on mechanosensing by
  glioma cells.
\newblock {\em New J. Phys.}, 16(7):075002, 2014.

\bibitem{Georges:2006qf}
P~C Georges, W~J Miller, D~F Meaney, E~S Sawyer, and
  P~A Janmey.
\newblock Matrices with compliance comparable to that of brain tissue select
  neuronal over glial growth in mixed cortical cultures.
\newblock {\em Biophys. J.}, 90(8):3012--8, 2006.

\bibitem{Flanagan:2002aa}
L~A Flanagan, Y-E Ju, B Marg, M Osterfield, and P~A Janmey.
\newblock Neurite branching on deformable substrates.
\newblock {\em Neuroreport}, 13(18):2411--5, 2002.

\bibitem{simon2013non}
M~Simon {\it et al}.
\newblock Non-invasive characterization of intracranial tumors by magnetic
  resonance elastography.
\newblock {\em New J. Phys.}, 15(8):085024, 2013.

\bibitem{Muthupillai:1995aa}
R~Muthupillai, D~J Lomas, P~J Rossman, J~F Greenleaf, A~Manduca, and R~L Ehman.
\newblock Magnetic resonance elastography by direct visualization of
  propagating acoustic strain waves.
\newblock {\em Science}, 269(5232):1854--7, 1995.

\bibitem{Manduca:2001aa}
A~Manduca, T~E Oliphant, M~A Dresner, J~L Mahowald, S~A Kruse, E~Amromin, J~P
  Felmlee, J~F Greenleaf, and R~L Ehman.
\newblock Magnetic resonance elastography: non-invasive mapping of tissue
  elasticity.
\newblock {\em Med. Image Anal.}, 5(4):237--54, 2001.

\bibitem{Katira:2012oq}
P Katira, M~H Zaman, and R~T Bonnecaze.
\newblock How changes in cell mechanical properties induce cancerous behavior.
\newblock {\em Phys. Rev. Lett.}, 108(2):028103, 2012.

\bibitem{Ven:2013aa}
A~L van~de Ven, B Abdollahi, C~J Martinez, L~A Burey,
  M~D Landis, J~C Chang, M Ferrari, and H~B Frieboes.
\newblock Modeling of nanotherapeutics delivery based on tumor perfusion.
\newblock {\em New J. Phys.}, 15:55004, 2013.

\bibitem{Kamoun:2010aa}
W~S Kamoun, S-S Chae, D~A Lacorre, J~A Tyrrell, M
  Mitre, M~A Gillissen, D Fukumura, R~K Jain, and L~L Munn.
\newblock Simultaneous measurement of RBC velocity, flux, hematocrit and shear
  rate in vascular networks.
\newblock {\em Nat. Methods}, 7(8):655--60, 2010.

\end{thebibliography}
\end{document}